\documentclass[11pt]{article}
\usepackage{hyperref}
\usepackage{amssymb}
\usepackage{graphicx}
\usepackage{float}
\usepackage{slashed}
\usepackage{amsmath}
\usepackage{tabularx}
\usepackage{breqn}
\usepackage{authblk}
\usepackage{array}
\usepackage{dsfont}
\usepackage{cite}
\usepackage{physics}
\usepackage[section]{placeins}
\usepackage[a4paper, total={6.8in, 10in}]{geometry}
\numberwithin{equation}{section}
\newcommand{\be}{\begin{equation}}
\newcommand{\ee}{\end{equation}}
\newcommand{\bea}{\begin{eqnarray}}
\newcommand{\eea}{\end{eqnarray}}
\newcommand{\ba}{\begin{aligned}}
\newcommand{\ea}{\end{aligned}}
\begin{document}
\title{Newtonian cosmology and Evolution of $\kappa$-deformed universe}

\author{E. Harikumar \thanks{eharikumar@uohyd.ac.in}, Harsha Sreekumar \thanks{harshasreekumark@gmail.com} and Suman Kumar Panja
\thanks{sumanpanja19@gmail.com}}
\affil{School of Physics, University of Hyderabad, \\Central University P.O, Hyderabad-500046, Telangana, India}

\maketitle
\begin{abstract} Considering space--time to be non-commutative, we study the evolution of the universe employing the approach of Newtonian cosmology. Generalizing the conservation of energy and the first law of thermodynamics to $\kappa$-deformed space--time, we derive the modified Friedmann equations, valid up to the first order, in the deformation parameter. Analyzing these deformed equations, we derive the time evolution of the scale factor in cases of radiation-dominated, matter-dominated, and vacuum (energy)-dominated universes. We show that the rate of change of the scale factor in all three situations is modified by the non-commutativity of space--time, and this rate depends on the sign of the deformation parameter, indicating a possible explanation for the observed Hubble tension. We undertake this investigation for two different realizations of non-commutative space--time coordinates. In both cases, we also argue for the existence of bounce in the evolution of the universe.
\end{abstract}


\section{Introduction}
Understanding the origin and evolution of the universe is one of the most intriguing issues in physics. The framework for addressing this issue is general relativity. The Friedmann equation, derived using general relativity, connects the universe's expansion rate to its energy density. Various scenarios were analyzed using the Friedmann equation.  Two basic ingredients in these studies are (i) the assumption of conservation of total energy and (ii) that changes in the volume and pressure of the universe are connected by the first law of thermodynamics. Interestingly, it was known in the very early days of cosmology that the Friedmann equation could be derived within the Newtonian framework itself  \cite{milne1,milne2,mcrea1,mcrea2,mcrea3,thatcher}, and many of the results derived using general relativity can be derived in this approach. In the Newtonian cosmology approach, one assumes the universe to be a sphere of mass $M$ and radius $R$ while a distant accelerating galaxy of mass $m$ is on the surface of this sphere. The force of attraction between the (rest of the) universe and the galaxy is the inverse square of the distance of separation, with the total mass of the universe concentrated at the origin. In  \cite{milne1,milne2,mcrea1,mcrea2,mcrea3}, for the small-sized universe, using co-moving coordinates, the Friedmann equation was derived using this approach. Using the conservation of the total non-relativistic energy and the first law of thermodynamics, the Friedmann equation was obtained  \cite{thatcher}, and variations of this scheme were developed and studied  \cite{Jordan1,Jordan2}.

In recent times, this approach is used to study the implication of quantum corrections to the Newtonian potential on the Friedmann equation and, thus, on the evolution of the universe \cite{Bargueno, Sunandan}. The quantum corrections to the Newtonian potential calculated using an effective field theory approach were used to set up the Friedmann equation in  \cite{Bargueno, Sunandan}. In these works, the variation in scale factor with time was used to investigate the evolution of the universe. The construction of the Friedmann equation, incorporating a quantum correction to the Newtonian potential, and analyzing the time evolution of the scale factor provides valuable insights  \cite{Bargueno, Sunandan}. These studies also help us achieve a better understanding of quantum corrections, and these can lead to clues for quantum gravity.

One of the most serious concerns in Physics is to understand the nature of gravity at the quantum regime. Several approaches such as string theory, loop quantum gravity, causal sets, dynamical triangulation, and non-commutative geometry  \cite{connes, douglas1,douglas2, rov, sorkin, Glikman, dop1,dop2, am,madore,seiberg-witten} are being pursued in this endeavor. One characteristic feature of microscopic gravity brought out by these approaches is the existence of a fundamental length scale  \cite{Glikman, dop1,dop2}. Non-commutative geometry inherently encodes a length scale and, thus, provides a testing ground to build quantum gravity models. Thus, it is of paramount interest to study how non-commutative space--time will affect the cosmological models. Since generalizing general relativity to non-commutative space--time and analyzing such models is harder  \cite{wess1,wess2,chaichian1,chaichian2,rivell,bal}, it is useful to adopt Newtonian cosmology to include non-commutative modifications and analyze the same. The effect of non-commutativity also changes the Newtonian potential  \cite{madfr,guha,hari2}. In  \cite{madfr}, it was shown that the Newtonian potential is modified in the power of radial distance, in the presence of extra dimensions, due to the non-commutativity of space--time (Doplicher--Fredenhagen--Roberts space--time). In  \cite{guha,hari2}, modifications to both kinetic and potential terms due to the non-commutativity of space--time were considered. Thus, it is interesting to study the effect of non-commutativity in the evolution of the universe within the Newtonian framework, and we take up this study here.

Various models on different non-commutative space--times and their implications have been studied extensively over the past few decades  \cite{douglas1,douglas2, dop1,dop2,wess1,chaichian1,chaichian2,kappa1,dimitrijevic,dasz,mel1, carlson,amorim}. Moyal space--time is a well-explored non-commutative space--time in recent times  \cite{douglas1,douglas2} whose coordinates satisfy the commutation relation
\be
 [\hat{x}_{\mu},\hat{x}_{\nu}]=i\theta_{\mu\nu},\nonumber
\ee
where $\theta_{\mu\nu}$ is a constant anti-symmetric tensor having the dimension of $(length)^2$. Moyal space--time violates the Lorentz symmetry, and its symmetry algebra is realized using Hopf algebra  \cite{wess1,chaichian1,chaichian2}. The $\kappa$-deformed space--time is another type of non-commutative space--time equipped with a Lie algebraic type space--time commutation relation. The corresponding symmetry algebra was also defined using Hopf algebra  \cite{kappa1,dimitrijevic,dasz,mel1}. In $\kappa$-space--time, space coordinates commute among themselves, but the commutation relation between time coordinates and space coordinates vary with space coordinates, i.e., they satisfy
\be
[\hat{x}^i,\hat{x}^j]=0,~~~[\hat{x}^0, \hat{x}^i]=ia\hat{x}^i,~~~a=\frac{1}{\kappa}. \label{ksp-1}
\ee
where $a$ is the deformation parameter having a dimension of $\textit{length}$. Since quantum theories of non-commutative space--time inherently include the fundamental length parameter (along with nonlocality and nonlinearity), it is of intrinsic interest to investigate physical phenomena in the presence of the fundamental length scale.

Various aspects of non-commutative gravity and physics on space--time have been studied in recent times. Different aspects of black hole physics in non-commutative space--time were analyzed in  \cite{gupta1,gupta2,gupta3,gupta4}. The correction to Bekenstein--Hawking entropy in non-commutative space-time was studied in  \cite{gupta5}. Different features of non-commutative field theoretical models in FLRW space--time were investigated in  \cite{paolo,jain}. Using squashed fuzzy sphere, the super Chandrasekhar limit for a white dwarf mass was calculated in  \cite{bani}. In~\cite{ohl}, using non-commutative differential calculus, Einstein's equation in a non-commutative setting was derived and analyzed. The analysis of cosmological models in non-commutative space--time provides us with the testing ground to study the effects of quantum gravity and may lead to testable signatures. Such studies can also lead to clues for further progress in the development of quantum gravity.

In this paper, we derive the Friedmann equations in $\kappa$-deformed space--time using Newtonian cosmology and analyze their implications. We start with the non-relativistic limit of the dispersion relation for a particle (galaxy) moving under the influence of the gravitational potential of the rest of the universe in the $\kappa$-deformed space--time. From the conservation of this energy, valid up to the first order in non-commutative parameter-$a$ and the first law of thermodynamics, we derive the Friedmann equations, valid up to the first order in $a$. Using these, we obtain the time evolution of the scale factor for the radiation-dominant, matter-dominant, and vacuum-energy-dominant universe. This analysis is performed for two different realizations of $\kappa$-deformed space--time coordinates and their functions.

The non-commutativity of space--time, introduced to study quantum gravity effects, is expected only at very high energies/extremely short distances of the order of the Planck scale. However, it is natural to anticipate that the existence of such a non-commutative structure leaves its signatures at low energy/large length scales. In this study, we are looking for possible modifications in the Newtonian model of cosmology. Since the deformation parameter $a$ is expected to be of the order of Planck length, we consider corrections valid up to the first order in $a$ here. We show that the rate of change of the scale factor is affected by the deformation of space--time. We also have derived a bound on the deformation parameter $a$, which is smaller than the Planck length.
 
This paper is organized in the following manner. Section \ref{s2} gives a brief description of the $\kappa$-deformed space--time and different realizations of the $\kappa$-space--time we use in our analysis. In Section \ref{s3}, we obtain the non-commutative corrections to the Friedmann equations up to the first order in deformation parameter $a$. In Section \ref{s3.1}, we find the corrected Friedmann equations in $\kappa$-space--time for $\varphi$ realization, and then, in Section \ref{s3.2}, we derive the same for $\alpha$, $\beta$, and $\gamma$ realization. In Section \ref{s4}, we study the evolution of the Newtonian universe in $\kappa$-space--time for these two different realizations. In Section \ref{s4.1}, we obtain the time development of the scale factor for radiation-dominated, matter/dust-dominated, and vacuum (energy)-dominated universe in $\kappa$-space--time for $\varphi$ realization. Next, in Section \ref{s4.2}, we derive the relation between the scale factor and time for $\alpha$, $\beta$, and $\gamma$ realization. Furthermore, we plot the variation of the scale factor with time for radiation-dominated, matter-dominated, and vacuum-energy-dominated universes and analyze these. In Section \ref{s5}, we give concluding remarks.

\section{Kappa Deformed Space--Time}\label{s2}

In this section, we summarize the essential details of the $\kappa$-deformed space--time that are needed for our purposes. $\kappa$-deformed space--time is a Lie-algebraic type non-commutative space--time, whose space--time coordinates satisfy Equation~(\ref{ksp-1}). The field theoretical models on the $\kappa$-space--time were constructed using star product formalism, where the usual notion of the pointwise product between the coordinates (and their functions) is replaced with the star product, which is invariant under the $\kappa$-Poincare algebra  \cite{dimitrijevic,dasz}. Alternatively, one can represent the non-commutative coordinates and their functions in terms of functions of commutative coordinates and their derivatives using realizations \cite{mel1,mel2}. It was shown that the realization approach is equivalent to star product formalism in $\kappa$-space--time  \cite{mel3}. For this study, we use the realization approach. We now discuss two different realizations of the $\kappa$-space--time  \cite{mel1, hari1a,hari1b} coordinates and their functions.

\subsection{$\varphi$-Realization} \label{s2.1}

The $\kappa$-deformed coordinate $\hat{x}_{\mu}$ is written in terms of the commutative coordinate $x_{\mu}$ and its derivatives $\partial_{\mu}$ as  \cite{mel1}
\begin{equation}\label{ksp-2}
\begin{split}
 \hat{x}_0=&x_0\psi(A)+iax_j\partial_j\gamma(A)\\
 \hat{x}_i=&x_i\varphi(A),
\end{split}
\end{equation}
where $A=ia\partial_0=-ap_{0}$, and $\psi$, $\gamma$, and $\varphi$ are functions of $A$, satisfying condition
\begin{equation}\label{ksp-3}
 \psi(0)=1,~\varphi(0)=1.
\end{equation}   
{By} substituting Equation~(\ref{ksp-2}) in Equation~(\ref{ksp-1}), we obtain:
\begin{equation}\label{ksp-4}
 \frac{\varphi'(A)}{\varphi(A)}\psi(A)=\gamma(A)-1,
\end{equation}
with $\varphi^{\prime}=\frac{d\varphi}{dA}$. Two possible realizations of $\psi(A)$ are $\psi(A)=1$ and $\psi(A)=1+2A$  \cite{mel1}. Now onwards, we choose $\psi(A)=1$. Thus, Equations (\ref{ksp-2}) and  (\ref{ksp-4}) become
\begin{equation}\label{ksp-5}
\begin{split}
 \hat{x}_0=&x_0+iax_j\partial_j\gamma(A)\\
 \hat{x}_i=&x_i\varphi(A),
\end{split}
\end{equation}
and
\begin{equation}\label{ksp-6}
 \frac{\varphi'(A)}{\varphi(A)}=\gamma(A)-1,
\end{equation} 
where the 
allowed choices of $\varphi$ are $e^{-A}, e^{-\frac{A}{2}}, 1, \frac{A}{e^A-1}$, etc.  \cite{mel1}. In  \cite{mel1}, it was shown that different choices of $\varphi$ correspond to different choices of ordering.
For example, $\varphi=e^{-A}$ leads to left ordering, $\varphi=1$ leads to right ordering, and $\varphi=\frac{A}{e^{A}-1}$ is for symmetric ordering. For this realization, the generic form of the free particle dispersion 
 relation is  \cite{hari2,mel1}
\be
\frac{4}{a^2}\sinh^2 \bigg(\frac{A}{2}\bigg) -p_ip_i \frac{e^{-A}}{\varphi^2(A)}-m^2c^2 +\frac{a^2}{4}\left[\frac{4}{a^2}\sinh^2\bigg(\frac{A}{2}\bigg) -p_ip_i \frac{e^{-A}}{\varphi^2(A)}\right]^2=0,\label{disp}
\ee
where $p_i$ are the momentum components corresponding to the commutative coordinates.

\subsection{$\alpha$, $\beta$, and $\gamma$-Realization}\label{s2.2}

Here, we starts with  \cite{hari1a,hari1b} 
\begin{equation}\label{ksp-7}
 \hat{x}_{\mu}=x_{\nu}\varphi^{\nu}_{\mu},
\end{equation}
where 
\be
 \varphi^{\nu}_{\mu}=\delta^{\nu}_{\mu}\Big(1+\alpha(a\cdot p)\Big)+\beta a^{\nu}p_{\mu}+\gamma p^{\nu}a_{\mu}, \label{ksp-7a}
\ee
and $\alpha$, $\beta$, and $\gamma$ are constants. Generalizing the commutation relation between the phase-space coordinates to corresponding non-commutative variables (i.e., $[\hat{x}_{\mu},\hat{p}_{\nu}]=i\hbar \hat{g}_{\mu \nu}$), we obtain the canonical momentum operator as  \cite{hari1a,hari1b} $\hat{p}_{\mu}= g_{\alpha \beta}(\hat{y})p^{\alpha}\varphi^{\beta}_{\mu}(p)$. In contrast to the realization discussed in the previous subsection, this realization maps the non-commutative phase-space variables into commutative phase-space variables with a linear dependence on the deformation parameter. This realization takes the form of:
 \bea
 \hat{x}^{\mu} &=& x^{\mu} + \alpha x^{\mu}(a \cdot p)+\beta(a \cdot x)p^{\mu}+\gamma a^{\mu}(x \cdot p) \\ \label{ksp-8}
 \hat{p}^{\mu} &=& p^{\mu} + (\alpha +\beta)(a \cdot p)p^{\mu}+\gamma a^{\mu}(p \cdot p). \label{ksp-9}
 \eea
{In} the above, $a \cdot p=a_{\mu}p^{\mu}$, $a \cdot x=a_{\mu}x^{\mu}$, $x \cdot p=x_{\mu}p^{\mu}$, and here, $\mu=0, 1, 2, 3$. These $\kappa$-deformed space--time coordinates, $\hat{x}^{\mu}$ and momentum, $\hat{p}^{\mu}$ satisfy the following Poisson brackets relations:
 \bea
 \left\lbrace \hat{x}^{\mu},\hat{x}^{\nu} \right\rbrace &=& a^{\mu}\hat{x}^{\nu}-a^{\nu}\hat{x}^{\mu} \\ \label{ksp-10}
 \left\lbrace \hat{p}^{\mu},\hat{p}^{\nu} \right\rbrace &=& 0 \\ \label{ksp-11}
 \left\lbrace \hat{p}^{\mu},\hat{x}^{\nu} \right\rbrace &=& \eta^{\mu \nu}\Big( 1+s(a \cdot p)+(s+2)a^{\mu}p^{\nu}+(s+1)a^{\nu}p^{\mu} \Big) \label{ksp-12}
 \eea
 where $s=2\alpha + \beta$. $a^{\mu}=(a,0)$ is the deformation parameter. $\alpha$, $\beta$, and $\gamma$ are real constants with $\gamma=1+\alpha$. In order to avoid the Hamiltonian/Lagrangian for simple models with explicit time dependence and to obtain the proper commutative limit (deformation parameter, \mbox{$a$ $\rightarrow$ 0}), we set $\beta=0$ (for details, see discussion in  \cite{vishnu1,guha}).

\section{Modified Friedmann Equations from Newtonian Dynamics}\label{s3}

In this section, we derive the non-commutative corrections to the Friedmann equations from Newtonian dynamics using the approach of  \cite{thatcher, Jordan1,Jordan2, Bargueno, Sunandan}. Using these, we analyze the expansion of the universe in the non-commutative setting. We derive the modified Friedmann equations for two different realizations  \cite{mel1,hari1a,hari1b}, as follows.

\subsection{$\varphi$-Realization} \label{s3.1}

For this study, we chose $\varphi=e^{-A}$. From the dispersion relation given in Equation~(\ref{disp}), we find the Hamiltonian in the non-relativistic limit, valid up to the first order in $a$, to be  \cite{hari2}:
\be
\hat{H}=\frac{(1-amc)}{2m}~ {\vec P}\cdot {\vec P},\label{kham}
\ee 
where ${\vec P}\cdot {\vec P}=\sum_{i} P_{i}P_{i},$ and ${\vec P}$ is the three vector momentum. Adding appropriate potential to the above Hamiltonian, we obtain the Hamiltonian describing the motion of a distant galaxy of mass $m$ at the surface of the sphere of radius $R$ with mass $M$ of the rest of the universe, uniformly distributed inside the sphere(and hence, we can take it to be concentrated at the center) to be:
\be
\hat{H}=\frac{P_{R}^2}{2m}(1-amc)-\frac{GMm}{R}(1-ap_{0}), \label{Fried-1}
\ee
valid up to the first order in $a$. Here, we have used $\hat{R}=R(1+ap_{0})$ \footnote{Using 
 \mbox{$\varphi=e^{-A}=e^{ap_{0}}$}, from Equation~(\ref{ksp-2}), we find $\hat{R}=\sqrt{\hat{x}_i\hat{x}^{i}}=\sqrt{x_ix^{i}(1+2ap_{0})}=R(1+ap_{0})$ and, hence, $-\frac{GMm}{\hat{R}}=-\frac{GMm}{R}(1-ap_{0})+O(a^2)$}, where $p_{0}=E$ is the energy of the commutative system, i.e., $p_{0}=E$ is the energy we find from Equation~(\ref{disp}) in the limit $a \rightarrow 0$. Note that we have considered correction terms only up to the first order in $a$. In the limit $a$ $\rightarrow$ 0, we retrieve the Hamiltonian for Newton's problem in the commutative space--time from the above equation. From the above equation, we find $P_{R}=m\frac{dR}{dt}(1+amc)$. Using this $P_{R}$ in the Hamiltonian in Equation~(\ref{Fried-1}), we find the total energy of the system valid up to the first order in $a$ to be:
 \be
E=\frac{1}{2}m\left(\frac{dR}{dt} \right)^{2}(1+amc)-\frac{GMm}{R}(1-ap_{0}), \label{Fried-2}
 \ee
where the first term represents the kinetic energy and the second term is the potential energy of the galaxy of mass $m$. This energy expression can be rewritten as
\be
E=\frac{1}{2}m\left(\frac{dR}{dt} \right)^{2}(1+aA)-\frac{GMm}{R}(1+aB), \label{Fried-2a}
 \ee
where $A$ and $B$ are constants that depend on $m$ and $p_{0}$.
The mass of the universe is given~by
\be
M=\frac{4}{3}\pi \hat{R}^{3}\rho \label{Fried-3}
\ee
where $\rho$ is the energy density  \cite{Jordan1,Jordan2, Bargueno, Sunandan} of the universe and $\hat{R}=R(1+ap_{0})$, obtained using Equation~(\ref{ksp-2}), is the radius of the universe. Next, we substitute the Equation~(\ref{Fried-3}) into Equation~(\ref{Fried-2}) and find:
\be
\frac{2E}{mR^{2}}= H^{2}(1+amc)-\frac{8\pi G \rho}{3}(1+2ap_{0}) \label{Fried-4}
\ee
where $H=\frac{1}{R}\frac{d R}{dt}$ is the Hubble parameter. Note here that we have considered modifications valid up to the first order in $a$. The above equation can be rewritten as:
\be
H^2(1+amc)-\frac{8\pi G \rho}{3}(1+2ap_{0})= -\frac{K\hat{R^2}(t_{1})}{R^{2}}, \label{Fried-5}
\ee
where $\hat{R}(t_{1})$ is the radius of the universe at time $t_{1}$, and $K=-\frac{2E}{m\hat{R^{2}}(t_{1})}$ is a constant that takes values of 1, 0, or $-$1, depending on  $E$ being negative, zero, or positive.
This Equation~(\ref{Fried-5}) is the first modified Friedmann equation having non-commutative correction terms valid up to the first order in the deformation parameter $a$. In the commutative limit, i.e., $a$ $\rightarrow$ 0, we retrieve the usual Friedmann equation  \cite{Jordan1,Jordan2, Bargueno}.
Now we find the second deformed Friedmann equation. For this, we consider the first law of thermodynamics for the expanding, homogeneous, and isotropic universe ($dQ=0$)
\be
dU+pdV=0, \label{Fried-6}
\ee
where $dQ$ is the differential increment of heat exchanged by the system (i.e., $dQ$ is the heat flow into or out of the volume), $dW = pdV$ is the differential element of work performed by the system, which is expanding, and $dU$ is the corresponding differential increase in the internal energy. Since $dQ=0$ for the homogeneous and isotropic universe, the first law of thermodynamics for the expanding universe in the $\kappa$-deformed space--time becomes:
\be
d\left(\rho \frac{4}{3}\pi \hat{R}^{3}\right)+ pd \left(\frac{4}{3}\pi \hat{R}^{3}\right)=0, \label{Fried-7}
\ee    
where $\hat{R}=R(1+ap_{0})$. Using this in the above equation, we find that the first law of thermodynamics for the homogeneous and isotropic universe is:
\be
d\left(\rho \frac{4}{3}\pi R^{3}\right)+ pd \left(\frac{4}{3}\pi R^{3}\right)=0, \label{Fried-8}
\ee
which is the same as that in the commutative space--time. Simplifying the above equation, we find the continuity equation
\be
 R\frac{d \rho}{d t}+3\left(p+\rho \right)\frac{dR }{d t}=0.\label{Fried-9}
\ee
{Next}, we multiply the first Friedmann equation by $R^{2}$ and take the derivative  with respect to $t$, which, after using Equation~(\ref{Fried-9}), we find the modified second Friedmann equation 
\be
\frac{d^{2}R }{d t^{2}}=-\frac{4 \pi G}{3}\left (3 p+\rho \right )R(1+2ap_{0}-amc)~. \label{Fried-10}
\ee
{Note} here that we have kept the modification to the second Friedmann equation only up to the first order in $a$. In the limit $a$ $\rightarrow$ 0, the above equation reduces to the well-known second Friedmann equation  \cite{Jordan1,Jordan2, Bargueno}. 

\subsection{$\alpha$, $\beta$ and $\gamma$-Realization} \label{s3.2}

In this case, we start with the Hamiltonian  \cite{guha}  constructed using the $\alpha$, $\beta$, and $\gamma$ realization~as:
\be
\hat{H}=\frac{P_{R}^2}{2m}(1+2\alpha a m)-\frac{GMm}{R}(1-\alpha ap_{0}), \label{Fried-11}
\ee
which represents the dynamics of a distant galaxy of mass $m$ at the surface of the sphere of radius $R$ with mass $M$ of the universe uniformly distributed inside the sphere. As before, $a$ is the deformation parameter, and $p_{0}$ is the energy level at which the non-commutativity of space--time comes into effect. Note here that we have included correction terms valid up to the first order in $a$. In the limit $a$ $\rightarrow$ 0, from the above equation, we retrieve the Hamiltonian for the commutative Newton problem. Next, from Equation~(\ref{Fried-11}), we find the total energy as:
 \be
E=\frac{1}{2}m\left(\frac{dR}{dt} \right)^{2}(1-2\alpha am)-\frac{GMm}{R}(1-\alpha ap_{0}), \label{Fried-12}
 \ee
which can be rewritten as:
\be
E=\frac{1}{2}m\left(\frac{dR}{dt} \right)^{2}(1+aC)-\frac{GMm}{R}(1+aD), \label{Fried-12a}
\ee
where $C$ and $D$ are constants, and they depend on $\alpha$, $m$, and $p_{0}$. Here, we observe that for both realizations, the non-commutative correction terms present in the expression of the conserved energy of the system (see Equations (\ref{Fried-2a}) and   (\ref{Fried-12a})) are of the same form. In these correction terms, $A$ and $B$ in Equation~(\ref{Fried-2a}) and $C$ and $D$ in Equation~(\ref{Fried-12a}) differ in numerical values depending on values of $m$, $p_{0}$, and $\alpha$.
Using $M=\frac{4}{3}\pi \hat{R}^{3} \rho$, the mass of the universe, where $\rho$ is the energy density of the universe, and $\hat{R}=R(1+\alpha a p_{0})$,  in Equation~(\ref{Fried-12}), we obtain:
\be
\frac{2E}{mR^{2}}= H^{2}(1-2 \alpha am)-\frac{8\pi G \rho}{3}(1+2ap_{0}). \label{Fried-14}
\ee
{Here}, $H=\frac{1}{R}\frac{d R}{dt}$ is the Hubble parameter. Thus, from the above equation, we find the deformed first Friedmann equation as:
\be
H^2(1-2 \alpha am)-\frac{8\pi G \rho}{3}(1+2 \alpha ap_{0})= -\frac{K\hat{R^2}(t_{1})}{R^{2}}, \label{Fried-15}
\ee
where $K=-\frac{2E}{m\hat{R^2}(t_{1})}$ is a constant. $K = 1,~ 0~$, and~$-1$ represent the closed, flat, and open universe, respectively  \cite{Mukanov,Steven}.
 In the commutative limit, i.e., $a$ $\rightarrow$ 0, we obtain the commutative Friedmann equation  \cite{Jordan1,Jordan2, Bargueno}. Note here that the $a$-dependent corrections are of the same form as in Equation~(\ref{Fried-5}). Thus, by following the same procedure discussed in Equations~(\ref{Fried-6})--(\ref{Fried-10}), we obtain the modified second Friedmann equation as:
\be
\frac{d^{2}R }{d t^{2}}=-\frac{4 \pi G}{3}\left (3 p+\rho \right )R(1+2\alpha ap_{0}+2 \alpha amc)~, \label{Fried-20}
\ee
{As} in the earlier case, here we also include corrections to the second Friedmann equation up to the first order in  $a$. In the limit $a$ $\rightarrow$ 0, the above equation reduces to the well-known second Friedmann equation  \cite{Jordan1,Jordan2, Bargueno}. 

 In the next section, we analyze the deformed Friedmann equations we have constructed in the $\kappa$-space--time.

\section{Non-Commutative Corrections to Scale Factor}\label{s4}

 In this section, we derive the non-commutative correction to the evolution of the universe in the Newtonian framework. We obtain modifications for scale factor ($A=\frac{R}{R_{0}}$) due to the non-commutativity of space--time. This is implemented using the Friedmann equation, Equation~(\ref{Fried-5}), and continuity equations obtained in Equation~(\ref{Fried-9}), for $\varphi$ realization. We analyze the scale factor for the radiation-dominated universe, dust-dominated universe, and vacuum (energy)-dominated universe. This is then repeated for $\alpha$, $\beta$. and $\gamma$ realization.

\subsection{For $\varphi$-Realization}\label{s4.1}
 
Here we have two independent equations out of the three equations, namely, the first  Friedmann equation, the second Friedmann equation, and the continuity equation. These equations connect three unknown parameters, $R(t)$, $\rho$, and $p$. We also have the relation between pressure and energy density, i.e., the equation of state
\be
p=\left(\gamma-1\right)\rho, \label{Ncnu-1}
\ee
where $\gamma$ is a constant \footnote{Note that in Sections \ref{s2} and   \ref{s3}, we have used $\gamma$ for representations of one particular realization of non-commutative functions (See Equations~(\ref{ksp-7a})--(\ref{ksp-9})). However, using $\gamma=1+\alpha$, we have replaced $\gamma$ in favor of $\alpha$. Thus, from now on, we use $\gamma$ representing a constant in Equation~(\ref{Ncnu-1})}, and it has a value of $\frac{4}{3}$ for the radiation-dominated universe~\cite{Steven}, $1$ for the matter-dominated universe, and $0$ for vacuum (energy)-dominated universe.
Using Equation~(\ref{Ncnu-1}) in Equation~(\ref{Fried-9}), we find:
 \be
\rho=\rho_{0}\left(\frac{R_{0}}{R}\right)^{3\gamma} \label{Ncnu-2}
\ee
where $\rho_{0} \equiv \rho(R_{0})$ is the energy density at the present time, with $R_{0}$ being the present radius of the universe.
Note here that Equation~(\ref{Ncnu-2}) does not contain any non-commutative correction. This is because the first law of thermodynamics for expanding the universe in Equation~(\ref{Fried-9}) does not receive any non-commutative corrections. Next, inserting Equations~(\ref{Ncnu-1}) and   (\ref{Ncnu-2}) into the modified first Friedmann equation Equation~(\ref{Fried-5}) with $K=0$ (i.e., for the flat universe)  \cite{Jordan1,Jordan2, Bargueno, Sunandan, Mukanov}, we find a corrected relation between the scale factor and time for the radiation dominated universe, the matter-dominated universe and vacuum (energy)-dominated~universe.

\subsubsection{Radiation $(\gamma=\frac{4}{3})$}\label{s4.1.1}

 Here we examine a spatially flat universe containing only radiation. In the radiation-dominated era (early universe), the matter density was negligible, and hence, one takes the universe to be flat. 
Setting $\gamma=\frac{4}{3}$ in Equation~(\ref{Ncnu-2}) and using it in Equation~(\ref{Fried-5}) for the radiation dominated universe, we obtain: 
\be
\left(\frac{d R}{d t}\right)^{2}=(1+2ap_{0}-amc)\frac{8 \pi G \rho_{0} R_{0}^2}{3}\Big(\frac{R_{0}}{R}\Big)^2. \label{Ncnu-3}
\ee
Here, we have considered corrections up to the first order in $a$. Next, we define the the scale factor $A=\frac{R}{R_{0}}$ and rewrite the above equation as:
\be
\left(\frac{d A}{d t}\right)^{2}=(1+2ap_{0}-amc)\frac{8 \pi G \rho_{0}}{3}\frac{1}{A^2} \label{Ncnu-4}
\ee
{Now} integrating the above equation, we obtain, for the radiation-dominated universe: 
\be
\frac{A^2}{2}=\pm \sqrt{(1+2ap_{0}-amc)\frac{8 \pi G \rho_{0}}{3}}~ t + C \label{Ncnu-5}
\ee
where $C$ is an integration constant. Setting the condition $R(t_{0})=R_{0}$  at present time $t_{0}$ gives 
\be
C = \frac{1}{2} \mp \sqrt{(1+2ap_{0}-amc)\frac{8 \pi G \rho_{0}}{3}}~ t_{0} \label{Ncnu-6}
\ee
{Substituting} this value for the constant in the Equation~(\ref{Ncnu-5}), we find that the scale factor $A(t)=\frac{R(t)}{R_{0}}$ satisfies
\be
\frac{A^2-1}{2}=\pm ~ \sqrt{(1+2ap_{0}-amc)\frac{8 \pi G \rho_{0}}{3}}~(t-t_{0}). \label{Ncnu-7}
\ee
{From} the above equation, we find the scale factor (in SI units)
\be
A_{\pm}=\sqrt{1 \pm 2\tau\Big(1-\frac{amc}{2\hbar}+\frac{ap_{0}}{\hbar}\Big) } , \label{Ncnu-8}
\ee
where $\tau=\sqrt{\frac{8 \pi G \rho_{0}}{3}}(t-t_{0})$. Here we have also considered correction up to the first order in $a$ only. In the limit $a$ $\rightarrow$ 0, the above equation reduces to the  commutative result, i.e.,
\be
A_{\pm}=\sqrt{1 \pm 2\tau}\label{Ncnu-9}
\ee
This solution is identical to the one obtained from general relativity  \cite{Steven}.

\begin{figure}[H]
 
\includegraphics[scale=.54]{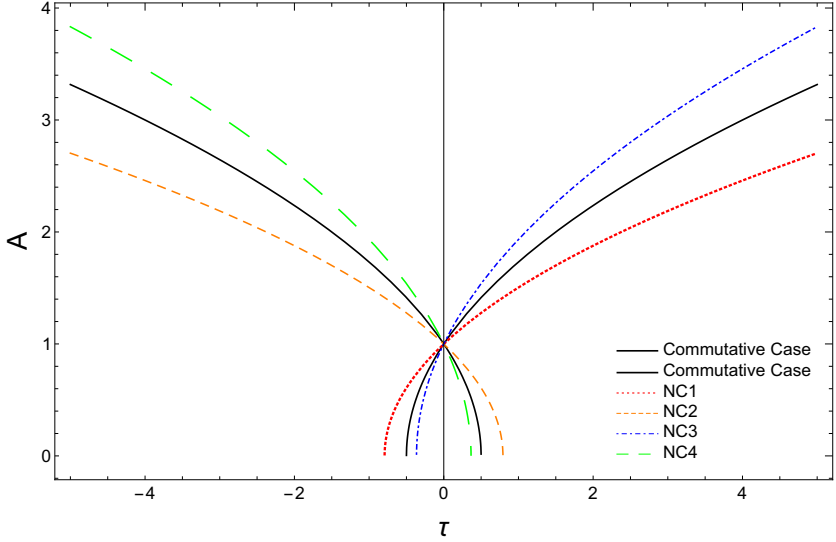}
\caption{Scale 
 factor versus $\tau$ for the radiation$-$dominated universe.}
\label{fig1}
\end{figure}

In Figure \ref{fig1}, we have plotted the variation of the scale factor $A_{\pm}$ for the flat ($K=0$), radiation-dominated universe for both the commutative case (Equation~(\ref{Ncnu-9})) and the modified case (Equation~(\ref{Ncnu-8})) with dimensionless quantity $\tau$. For this, we have taken $m$ to be the mass of the Andromeda galaxy, i.e., $m \sim 1.3 \times 10^{12}~M_{\odot}$ and $p_{0}$ is assumed to be equal to the mass of the extra-galactic black hole, i.e., $10^{6}~M_{\odot}$ (at this energy level, the non-commutative effect is assumed to be important). Here, $M_{\odot}$ is one solar mass. Since corrections due to the non-commutativity of space--time are expected to be very small, after inserting the above numerical values in the Equation~(\ref{Ncnu-8}), we find that the deformation parameter $a$ becomes bounded, i.e., $|a|\leq 10^{-85} m$. In Figure \ref{fig1}, it is important to note that for the positive value of deformation parameter $a$, we find two curves, NC1 (red dotted curve) and NC2 (orange dashed curve). Similarly, for a negative value of $a$, we find two curves, NC3 (blue dot-dashed curve) and NC4 (green dashed curve). For a positive $a$, curve (NC1) shows that the scale factor ($A_{+}$) is increasing slower than the commutative case at a future time (i.e.,$\tau > 0$) and scale factor $A_{-}$ (NC2-dashed orange curve) is decreasing slower than the commutative case at the past time (i.e., $\tau < 0$). This implies that for the positive value of deformation parameter $a$, the universe is contracting slower than the commutative case, and the expansion of the universe is also slower than the commutative case. For a negative $a$, the scale factor-$A_{+}$ (NC3) at a future time (i.e., $\tau > 0$) is increasing more rapidly than the commutative case for expanding the radiation-dominated universe, and for a negative $a$, the scale factor-$A_{-}$ (NC4) at the past time (i.e., $\tau < 0$) is decreasing faster than the commutative case. This implies that for a negative $a$, the contraction of the universe is faster compared to the commutative case, and the expansion rate of the universe is higher than the commutative case. Note here that at the present time, i.e., at $\tau=0$ ($t=t_{0})$, all these four curves intersect at $A=1$.
Note that the expansion rate of the universe  in the non-commutative case is different from the commutative case. The expansion rate is slower than in the commutative case when $a$ is positive, and 
it is faster when $a$ is negative, in comparison with the commutative case. In  \cite{hu1,hu2}, it is shown that there is a discrepancy in the value of the Hubble parameter obtained by different types of measurements. In one approach, $H_{0}$ is found to be $67.40 \pm 1.40$ km/s/Mpc   \cite{planc}, and in another, it is found to be $70.00 \pm 2.20$ km/s/Mpc \cite{ben} in the present time ($\tau=0$). This discrepancy is known as Hubble tension \cite{hu1,hu2}. In Figure~\ref{fig1}, we observe that the expansion rate of the universe is different depending on the sign of the deformation parameter $a$. Thus, one may be able to interpret the Hubble tension using the non-commutativity of space--time.

\subsubsection{Dust $(\gamma=1)$}

 In this subsection, we consider the matter-dominated universe, which is spatially flat. Thus ,we set $K=0$ and $\gamma=1$ in Equations (\ref{Fried-5}) and   (\ref{Ncnu-2}) and, using these two equations, we obtain:
 \be
\left(\frac{d R}{dt}\right)^{2}=(1+2ap_{0}-amc) \frac{8\pi G \rho_{0}R_{0}^2}{3}\Big(\frac{R_{_{0}}}{R}\Big)\label{Ncnu-10}
\ee
where we have taken correction terms up to the first order in deformation parameter $a$. Rewriting the above equation as:
\be
\left(\frac{d A}{d t}\right)^{2}=(1+2ap_{0}-amc) \frac{8\pi G \rho_{0}}{3}\Big(\frac{1}{A}\Big),\label{Ncnu-11}
\ee
and integrating,  we obtain:
\be
\frac{2}{3}A^{\frac{3}{2}}=\pm \sqrt{(1+2ap_{0}-amc)\frac{8\pi G \rho_{0}}{3}}~t + C \label{Ncnu-12}
\ee
where $C$ is an integration constant. Using the condition that $t=t_{0}$ ($t_{0}$ is the present time), $R=R_0$, and $A=1$ as before, we obtain:
\be
C= \frac{2}{3} \mp \sqrt{\frac{(1+2ap_{0}-amc)8\pi G \rho_{0}}{3}}~t_{0}. \label{Ncnu-13}
\ee
{Substituting} this in Equation~(\ref{Ncnu-12}), we find: 
\be
\frac{2}{3}(A^{\frac{3}{2}}-1)=\pm ~ \sqrt{(1+2ap_{0}-amc)\frac{8\pi G \rho_{0}}{3}}~(t-t_{0}). \label{Ncnu-14}
\ee
{After} simplifying the above equation, we find the scale factor (in SI unit)
\be
A_{\pm}=\Bigg(1\pm \frac{3}{2}\tau \Big(1-\frac{amc}{2\hbar}+\frac{ap_{0}}{\hbar}\Big)\Bigg)^{\frac{2}{3}}, \label{Ncnu-15}
\ee
where $\tau=\sqrt{\frac{8 \pi G \rho_{0}}{3}}(t-t_{0})$. The commutative solution is recovered by taking the limit $a \rightarrow 0$ in the above equation, which reads:
\be
A_{\pm}=\left(1\pm\frac{3}{2}\tau\right)^{\frac{2}{3}}. \label{Ncnu-16}
\ee
{As} in the radiation case, this solution is identical to the one obtained using general relativity~\cite{Steven}. 

\begin{figure}[H]
 \includegraphics[scale=.54]{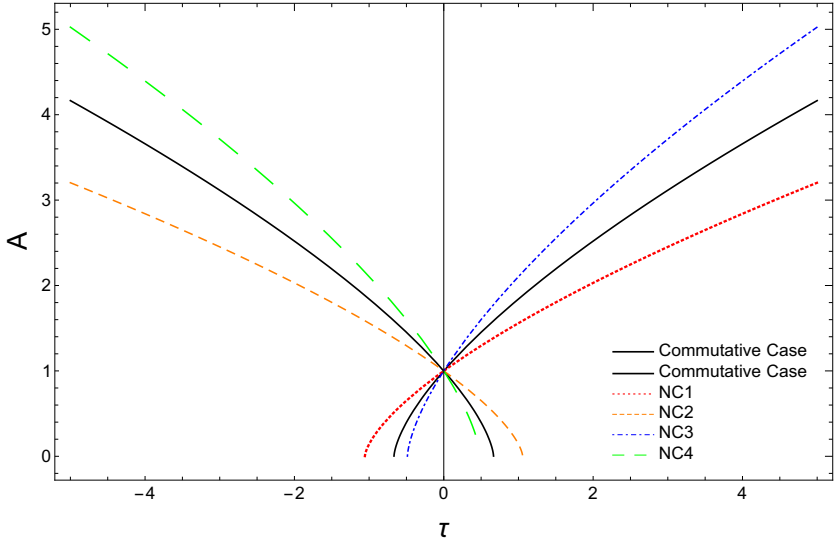}
\caption{Scale 
 factor $A$ versus $\tau$ in the dust (matter)$-$dominated universe.}
\label{fig2}
\end{figure}

In Figure \ref{fig2}, we have plotted the scale factor $A_{\pm}$ against $\tau$ for the matter-dominated universe for both the commutative case (Equation~(\ref{Ncnu-16})) and the non-commutative corrected case (Equation~(\ref{Ncnu-15})), with $\tau$. As in the radiation-dominated universe case, here also we take the same numerical values for $m$ (mass of Andromeda galaxy) and $p_{0}$ (obtained from black hole mass). Here also we obtain the bound for the deformation parameter $a$ as $|a|\leq 10^{-85} m$. From the plots in Figure \ref{fig2}, we find that in the case of the matter-dominated flat universe, the scale factor $A_{\pm}$  behaves similar to the spatially flat radiation-dominated case (discussed in Section~\ref{s4.1.1}) depending on the sign of deformation parameter $a$, with $\tau$.

\subsubsection{Vacuum $(\gamma=0)$}

\begin{figure}[H]
 
\includegraphics[scale=.54]{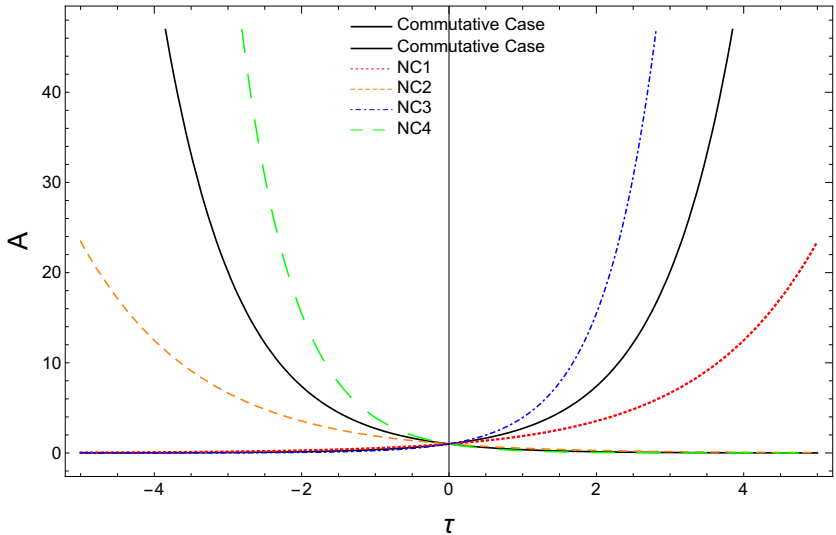}
 \caption{Scale 
 factor $A$ versus $\tau$ for the vacuum (energy)$-$dominated universe.} 
\label{fig3}
\end{figure}

 In this subsection, we study the vacuum-energy-dominated universe. For this, we use  Equations (\ref{Fried-5}) and   (\ref{Ncnu-2}) with $\gamma=0$ and find:
\be
\left(\frac{d R}{d t}\right)^{2}=(1+2ap_{0}-amc)\frac{8\pi G \rho_{0} R_{0}^2}{3}\Big(\frac{R}{R_{0}}\Big)^2 \label{Ncnu-17}
\ee
{Here} also we consider correction terms up to the first order in $a$. The above equation can be rewritten as:
\be
\left(\frac{d A}{d t}\right)^{2}=(1+2ap_{0}-amc) \frac{8\pi G \rho_{0}}{3}A^2, \label{Ncnu-18}
\ee
which, on integration, gives:
\be
ln(A)=\pm ~ \sqrt{(1+2ap_{0}-amc)\frac{8\pi G \rho_{0}}{3}}~t + C. \label{Ncnu-19}
\ee
{Here}, $C$ is an integration constant, which, by applying the conditions  $R(t_{0})=R_0$ and $A=1$, is fixed as
\be
C= \mp ~ \sqrt{\frac{(1+2ap_{0}-amc)8\pi G \rho_{0}}{3}}~t_{0}. \label{Ncnu-20}
\ee
{Now}, using this expression of $C$ in Equation~(\ref{Ncnu-19}), we find the scale factor (in SI unit) to~be:
\be
A_{\pm}=\exp \Bigg(\pm \tau \Big(1-\frac{amc}{2\hbar}+\frac{ap_{0}}{\hbar}\Big)\Bigg), \label{Ncnu-22}
\ee
where $\tau=\sqrt{\frac{8 \pi G \rho_{0}}{3}}(t-t_{0})$.  In the limit, $a \rightarrow 0$ from the above equation reduces to
\be
A_{\pm}=\exp \left(\pm \tau\right) \label{Ncnu-23}
\ee
which is the commutative expression of the scale factor for the vacuum (energy)-dominated universe  \cite{Bargueno, Sunandan}.
The scale factor $A$ vs. $\tau$ plots are illustrated in Figure \ref{fig3} for a spatially flat, vacuum (energy)-dominated universe. Here also we have taken the same numerical values for  $m$ (mass of Andromeda galaxy ) and $p_{0}$ as before. As in the earlier case, we find that the deformation parameter $a$ receives the same upper bound, i.e., $|a|\leq 10^{-85} m$. For a positive $a$, the curve (NC1-red dotted curve) shows that the scale factor ($A_{+}$) is increasing exponentially but at a slower rate than the commutative case in the future time (i.e., $\tau > 0$), and the scale factor $A_{-}$ (NC2-orange dashed curve) decreased exponentially but slower than the commutative case for the past time (i.e., $\tau < 0$). For a negative $a$, at a future time (i.e., $\tau > 0$), the variation of the scale factor-$A_{+}$ (NC3-blue dot-dashed curve) shows that the universe is expanding exponentially but more rapidly than the commutative case, and at the past time (i.e., $\tau < 0$), the variation of scale factor-$A_{-}$ (NC4-green dashed curve) shows that the universe is contracting exponentially but faster than the commutative case. Note here that in these four cases also, at the present time (i.e., $\tau=0$), the scale factor intersects at $A=1$.   

\subsection{For $\alpha$, $\beta$ and $\gamma$-Realization}\label{s4.2}

For this realization, we have two independent equations, namely, the first  Friedmann equation given in Equation~(\ref{Fried-15}) and the continuity equation (similar to Equation~(\ref{Fried-9}) adapted for this realization). As in the previous case, for $\alpha$, $\beta$, and $\gamma$ realization also, we find that the Friedmann equation is modified, and the continuity equation remains unchanged. Here also we have the same relation between pressure and energy density (See Equation~(\ref{Ncnu-1})), which leads to the same relation between $\rho$ and $\rho_{0}$ as given in Equation~(\ref{Ncnu-2}).
 
Note that the modified Friedmann's equations (\ref{Fried-5}) and   (\ref{Fried-15}) in the
non-commutative space--time are of the same form (see the discussion after Equation~(\ref{Fried-12a})). Thus, by following the same procedure discussed in the previous section (see  from \mbox{Equations~(\ref{Ncnu-3})--(\ref{Ncnu-7})}), we find the scale factor for the  radiation-dominated universe ($\gamma=\frac{4}{3}$)  as (in the SI unit):
\be
A_{\pm}=\sqrt{1 \pm 2\tau\Big(1+\frac{\alpha ap_{0}}{\hbar}+ \frac{\alpha amc}{\hbar})\Big) } , \label{Ncnu1-8}
\ee
where $\tau=\sqrt{\frac{8 \pi G \rho_{0}}{3}}(t-t_{0})$. In the limit, $a$ $\rightarrow$ 0, the above equation reduces to the  commutative result $A_{\pm}=\sqrt{1 \pm 2\tau}$ for the radiation-dominated universe. This solution also matches the one obtained from general relativity \cite{Steven}.

\begin{figure}[H]
 
\includegraphics[scale=.54]{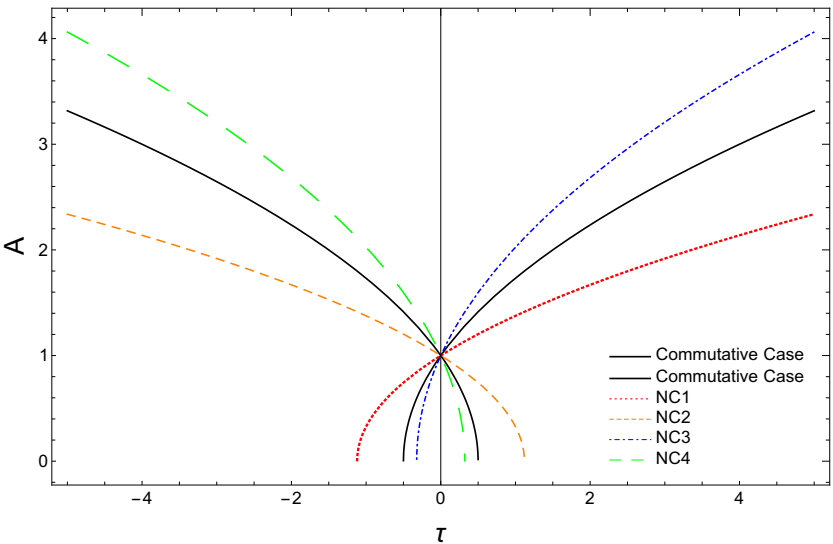}
\caption{Scale 
factor (A) versus $\tau$ for radiation$-$dominated universe.}
\label{fig4}
\end{figure}

In Figure \ref{fig4}, we have plotted the variation of the scale factor $A_{\pm}$ for the matter-dominated universe for both the commutative case and non-commutative corrected case (Equation~(\ref{Ncnu1-8})), with $\tau$. Taking the same numerical values for $m$ (mass of Andromeda galaxy) and $p_{0}$ (obtained from black hole mass) as before, and using $\alpha=-\frac{3}{4}$  \cite{vishnu1}, we find the bound for the deformation parameter $a$ as $|a|\leq 10^{-85} m$.  From the Figure \ref{fig4}, we find that for $\alpha$, $\beta$, and $\gamma$ realizations in the case of the radiation-dominated flat universe, the curves show similar behavior as in the spatially flat, radiation-dominated case (discussed in Section~\ref{s4.1.1}) depending on the sign of deformation parameter $a$.

\begin{figure}[H]
 
\includegraphics[scale=.54]{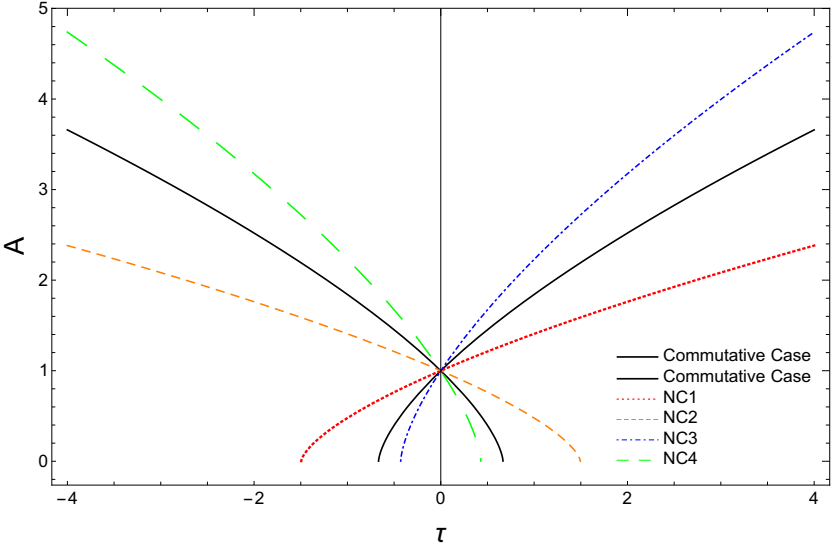}
\caption{Scale 
factor versus $\tau$ for the dust (matter)$-$dominated universe.}
\label{fig5}
\end{figure}

As in the earlier case (see from Equations~(\ref{Ncnu-10})--(\ref{Ncnu-15})), here we find the scale factor for the matter-dominated spatially flat universe ($\gamma=1$) to be
\be
A_{\pm}=\Bigg(1\pm \frac{3}{2}\tau \Big(1+\frac{\alpha ap_{0}}{\hbar}+ \frac{\alpha amc}{\hbar}\Big)\Bigg)^{\frac{2}{3}}, \label{Ncnu1-15}
\ee
where $\tau=\sqrt{\frac{8 \pi G \rho_{0}}{3}}(t-t_{0})$. The commutative solution can be recovered by taking the limit $a \rightarrow 0$ in the above equation and reads $A_{\pm}=\left(1\pm\frac{3}{2}\tau\right)^{\frac{2}{3}}$. As before, this solution is identical to the solution obtained using general relativity  \cite{Steven}. 
In Figure \ref{fig5}, we have shown the scale factor $A_{\pm}$, for the matter-dominated universe for both the commutative case and the non-commutative corrected case (Equation~(\ref{Ncnu1-15})), with $\tau$. Using the same numerical values for $m$, $p_{0}$, and $\alpha$ as before, we obtain the same bound for deformation parameter $a$ as $|a|\leq 10^{-85} m$. From Figure \ref{fig5}, we find that in the case of the matter-dominated, flat universe, scale factor $A_{\pm}$  behaves similarly to the spatially flat radiation-dominated case depending on the positive/negative value of $a$.
{For the vacuum (energy)-dominated universe ($\gamma=0$), we obtain (see Equations~(\ref{Ncnu-17})--(\ref{Ncnu-22}))}:
\be
A_{\pm}=\exp \Bigg(\pm \tau \Big(1+\frac{\alpha ap_{0}}{\hbar}+ \frac{\alpha amc}{\hbar}\Big)\Bigg), \label{Ncnu1-22}
\ee
where $\tau=\sqrt{\frac{8 \pi G \rho_{0}}{3}}(t-t_{0})$.  In the commutative limit, $a \rightarrow 0$, the above equation reduces to $
A_{\pm}=\exp \left(\pm \tau\right) $, which is the commutative expression of the scale factor for the vacuum universe  \cite{Sunandan, Steven}.

\begin{figure}[H]
 
\includegraphics[scale=.54]{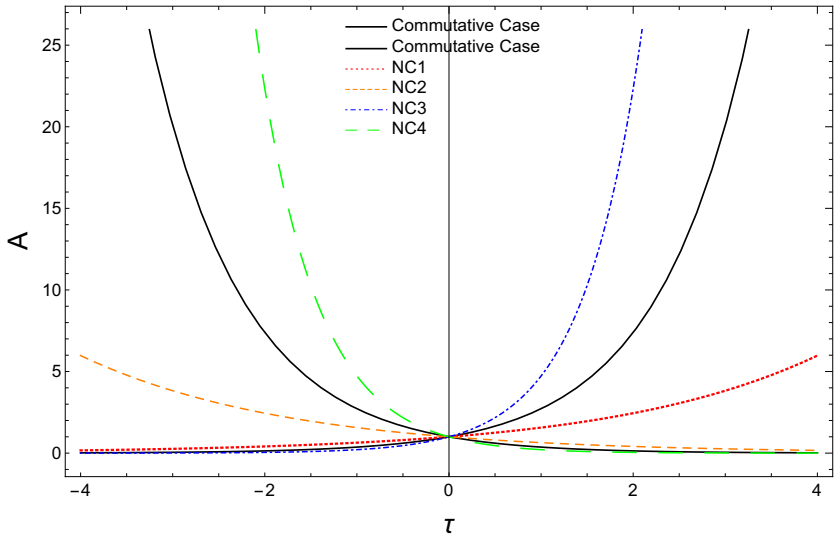}
 \caption{Scale
 factor (A) versus $\tau$ for vacuum (energy)$-$dominated universe.} 
\label{fig6}
\end{figure}

The plots of the scale factor $A$ vs. $\tau$ are given in Figure \ref{fig6} for a flat vacuum (energy)-dominated universe. As in the previous cases, we have taken the same numerical values for  $m$, $p_{0}$, and $\alpha$ and found that the deformation parameter $a$ receives the same bound, i.e., $|a|\leq 10^{-85} m$. In Figure \ref{fig6}, we observe that for positive $a$ at a future time (i.e., $\tau > 0$), the scale factor ($A_{+}$) is increasing exponentially (NC1-red dotted curve) but less rapidly than the commutative case, and at the past time (i,e., $\tau < 0$), the scale factor $A_{-}$ (NC2-dashed orange curve) is decreasing exponentially slower than the commutative case. For a negative $a$, at a future time ($\tau > 0$), the scale factor($A_{+}$) is increasing exponentially (NC3-blue dot-dashed curve) faster than the commutative case, and for a past time ($\tau < 0$), the scale factor (NC4-green dashed curve) is decreasing exponentially faster than the commutative case. As in the vacuum universe case for $\varphi$-realization, here also these two curves imply that for positive $a$, the rate of expansion of the universe is slower than the commutative case, and for negative $a$, the universe is expanding faster than the commutative case. Note here that at the present time, we do not observe any deviation in the scale factor $A_{\pm}$ for a flat vacuum universe.

\section{Conclusions}\label{s5}

In this paper, we have derived and analyzed solutions to the Friedmann equation in the $\kappa$-space--time for two different realizations  \cite{mel1,hari1a,hari1b}. In Newtonian cosmology, one assumes the universe as a sphere of mass M and any astrophysical object (distant galaxy) with mass m located at the sphere's surface. The astrophysical object feels gravitational force towards the center of the sphere (universe). This gravitational force is proportional to the mass contained in the universe and an inverse square to the distance between
the galaxy and the center of the universe. We have used the Newtonian cosmology framework for the analysis presented here. There are studies constructing non-commutative models of general relativity. These models are highly mathematical and, hence, applying them to cosmology requires rigorous calculation and may mask many features of the solutions. Thus, it is of immense interest to study cosmological solutions, taking into account the non-commutativity of space--time, using simpler approaches. In this study, we adopt Newtonian cosmology to investigate the modification to the Friedmann equations due to the non-commutativity of space--time and analyze their implications. This was performed by starting from the energy conservation of a particle (galaxy) of mass m, and using the first law of thermodynamics, we derived the Friedmann equations. We also derived the continuity equation from the first law of thermodynamics. We have incorporated correction terms valid up to the first order in the deformation parameter $a$. Using the modified Friedmann equation and continuity equation, we found the relation between the scale factor $A_{\pm}$ and $\tau$ (time) for the radiation-dominated universe, matter-dominated universe, and vacuum (energy)-dominated universe in the case of $\varphi$-realization and $\alpha$, $\beta$, and $\gamma$ realization, respectively. Using relations between the scale factor $A_{\pm}$ and $\tau$, we investigated the evolution of the Newtonian universe and plot scale factor against time for the radiation-dominated, dust (matter)-dominated, and vacuum (energy)-dominated cases. 

In  \cite{madfr}, the non-commutative corrections to the energy-momentum tensor were analyzed. Following this procedure, we found that in $\kappa$-deformed space--time, the density valid up to the first order in $a$ is $\hat{\rho}=\rho(1+ap_{0})$, where $p_{0}$ is the energy scale introduced in Equation~(\ref{Fried-1}). We note that under this change in the density (along with the corresponding change in the pressure), Equation 
(\ref{Fried-8}), Equation~(\ref{Fried-9}) for $\varphi$-realization, and the corresponding equations for $\alpha$, $\beta$, and $\gamma$ realizations do not change. Thus the result reported here remains valid even when the non-commutative corrections to density and pressure are~included.

We note that for both the realizations, the evolution of the universe in the radiation-dominated as well as the matter-dominated cases is similar(see the plots). This is expected as the deformed energy  expressions in Equations (\ref{Fried-2a}) and   (\ref{Fried-12a}) show that the non-commutative corrections in both cases are of the same generic form. Thus, it is clear that the actual evolution of the universe depends on the choice of the non-commutative parameters $a$ and $\alpha$. We observe that in all these cases ($\gamma=\frac{4}{3}, 1$, and $0$), the evolution of the universe is similar to the case of the quantum-corrected Newtonian universe  \cite{Bargueno, Sunandan}. 

We also observe that the evolution of the universe is different depending on the sign of the deformation parameter $a$. For positive values of $a$, contraction and expansion rates are slower than the commutative case, while for negative values of $a$, contraction and expansion rates are faster than the commutative case. This shows that different types of corrections due to quantum gravity effects can have contrasting effects on the Hubble parameter. It indicates that the observed tension in the Hubble parameter may be due to differences in the quantum gravity corrections seen in different measurements. We have shown that the expanding rate of the universe depends upon the sign of the deformation parameter $a$. This may also be relevant in explaining the difference in the Hubble constant given by Supernovae data  \cite{dhawan1,dhawan2} and Planck data  \cite{planc}.

In  \cite{hagh}, the neutrino flux was calculated in the Moyal space--time, and by comparing this with the observed neutrino flux from the Supernovae SN1987A, a bound on the non-commutative parameter $\theta^{\mu \nu}$ was obtained. The construction of gauge theoris in $\kappa$-deformed space--time is still in progress and, hence, such an analysis for $\kappa$-space--time is more complicated. In  \cite{Joby}, by considering the scalar perturbations in the Moyal space--time, the effect of non-commutativity on the CMB angular power spectrum was studied, and comparing with the CMB temperature fluctuations, a bound on the non-commutative parameter $\theta^{\mu\nu}$ was derived. These authors have also used data from other observations and calculated bounds on $\theta^{\mu\nu}$. Since scalar theory in $\kappa$-deformed space--time is well-studied, one can generalize the above analysis to $\kappa$-deformed space--time. Work along these lines is in progress.

Analyzing the evolution of the universe, we observe the existence of bounce. In analogy with quantum mechanics, one finds that Equation~(\ref{ksp-1}) leads to the uncertainty relation $\Delta \hat{x}_{0} \Delta \hat{x}_{i} \geq \frac{a}{2}|<\hat{x}_{i}>|$. Since Equation~(\ref{Ncnu-8}) implies that at $A=0$, the time $\tau=\mp\frac{1}{2}\Big(1+\frac{amc}{2\hbar}-\frac{ap_{0}}{\hbar}\Big)$, and this requires measuring $\tau$ to a very great accuracy (since $a \sim 10^{-85} m$), one will not be able to localize the universe below a certain length scale due to the above uncertainty relation. This prevents singularity in the case of the evolution of the universe. Thus, the non-commutativity of space--time avoids the singularity in the evolution of the Newtonian universe, indicating a bouncing universe.\\

\noindent {\bf Acknowledgments} S.K.P. thank UGC, India, for the support through the JRF scheme (id.191620059604).


\begin{thebibliography}{999}
\bibitem{milne1} McCrea, 
 W.H.; Milne, E.A. \textit{Newtonian Universe and the Curvature of Space}.    \textit{Quart. J. Math. Oxf.} \textbf{1934}, \emph{5}, 73.
 
 \bibitem{milne2} Milne, E.A. \textit{A Newtonian Expanding Universe}. \textit{Quart. J. Math. Oxf.} \textbf{1934}, \emph{5}, 64. 

\bibitem{mcrea1} McCrea, 
 W.H. \textit{On the significance of Newtonian cosmology}.   \textit{Astron. J.} \textbf{1955}, \emph{60}, 271.
 
 \bibitem{mcrea2}McCrea, W.H. \textit{Newtonian Cosmology}.  \textit{Nature} \textbf{1955}, \emph{175}, 466.
 \bibitem{mcrea3} Callan, C.; Dicke, R.H.; Peebles, P.J.E. \textit{Cosmology and Newtonian Mechanics }.  \textit{Am. J. Phys.} \textbf{1965}, \emph{33}, 105.

\bibitem{thatcher} Thatcher, A.R. \textit{Newtonian cosmology and Friedmann's equation}. \textit{Eur. J. Phys.} \textbf{1982}, \emph{3}, 202.

\bibitem{Jordan1} Jordan, 
 T.F.  \textit{Cosmology calculations almost without general relativity}. \textit{Am. J. Phys.}  \textbf{2005}. \emph{73}, 653.
 \bibitem{Jordan2}Tipler, F.J.  \textit{Rigorous Newtonian cosmology}. \textit{Am. J. Phys.} \textbf{1996}, \emph{64}, 1311.

\bibitem{Bargueno} Bargueño, P.; Medina, S.B.; Nowakowski, M.; Batic, D. \textit{Newtonian cosmology with a quantum bounce}. \textit{Eur. Phys. J.} \textbf{2016}, \emph{76}, 543. 

\bibitem{Sunandan} Mandal, R.; Gangopadhyay, S.; Lahiri, A. \textit{Newtonian cosmology from quantum corrected Newtonian potential}. \textit{Phys. Lett.} \textbf{2023}, \emph{839}, 137807.

\bibitem{connes} Connes, A. \textit{Non-Commutative Geometry}; Academic Press: London, UK, 1994.

\bibitem{douglas1} Douglas, 
 M.R.; Nekrasov, N.A. \textit{Noncommutative field theory}. \textit{Rev. Mod. Phys.} \textbf{2001}, \emph{73}, 977.
 \bibitem{douglas2}Szabo  R.J. \textit{Quantum field theory on noncommutative spaces}. \textit{Phys. Rep.} \textbf{2003}, \emph{378}, 207.

\bibitem{rov} Rovelli, C. \textit{Quantum Gravity}; Cambridge University Press: Cambridge, UK, 2004. 
\bibitem{sorkin} Bombelli, L.; Lee, J.; Meyer, D.; Sorkin, R.D.
\textit{Space-time as a causal set}. \textit{Phys. Rev. Lett.} \textbf{1987}, \emph{59}, 521. 

\bibitem{Glikman}
Kowalski-Glikman, J. \textit{Introduction to Doubly Special Relativity}. \textit{Lect. Notes. Phys.} \textbf{2005}, \emph{669}, 131.

\bibitem{dop1} Doplicher, 
 S.; Fredenhagen, K.; Roberts, J.E. \textit{Spacetime quantization induced by classical gravity}. \textit{Phys. Lett.} \textbf{1994}, \emph{331}, 29.
 \bibitem{dop2}Doplicher, S.; Fredenhagen, K.; Roberts, J.E. \textit{The quantum structure of spacetime at the Planck scale and quantum fields}.  \textit{Commun. Math. Phys. } \textbf{1995}, \emph{172}, 187.

\bibitem{am}
Ambjorn, J.; Jurkiewicz, J.; Loll, R. \textit{Emergence of a 4D World from Causal Quantum Gravity}. \textit{Phys. Rev. Lett.} \textbf{2004}, \emph{93}, 131301.

\bibitem{madore}Madore, J. \emph{An Introduction to Non-Commutative Differential Geometry and Its Applications}; Cambridge University Press: Cambridge, UK, 1995.

\bibitem{seiberg-witten} Seiberg, N.; Witten, E. \textit{String theory and noncommutative geometry}. \textit{J. High Energy Phys.} \textbf{1999}, \emph{9909}, 032.

\bibitem{wess1} Aschieri, P.; Blohmann, C.; Dimitrijevic, M.; Meyer, F.; Schupp, P.; Wess, J. \textit{A gravity theory on noncommutative spaces}. \textit{Class. Quantum Grav.} \textbf{2005}, \emph{22}, 3511.

\bibitem{wess2} Aschieri, P.A.; Dimitrijevic, M.; Meyar, F.; Wess, J. \textit{Noncommutative geometry and gravity}. \textit{Class. Quantum Grav.} \textbf{2006}, \emph{23}, 1883.

\bibitem{chaichian1} 
Chaichian
, M.; Kulish, P.; Nishijima, K.; Tureanu, A. \textit{On a Lorentz-invariant interpretation of noncommutative space–time and its implications on noncommutative QFT}. \textit{Phys. Lett.} \textbf{2004}, \emph{604}, 98.
\bibitem{chaichian2}Chaichian, M.; Tureanu, A.; Zet, G. \textit{Corrections to Schwarzschild solution in noncommutative gauge theory of gravity}. \textit{Phys. Lett. } \textbf{2008}, \emph{660}, 573.

\bibitem{rivell} Harikumar, E.; Rivelles, V.O. \textit{Noncommutative gravity}. \textit{Class. Quantum Grav.} \textbf{2006}, \emph{23}, 7551.

\bibitem{bal} Balachandran, A.P.; Govindarajan, T.R.; Gupta, K.S.; Kurkcuoglu, S. \textit{Noncommutative two-dimensional gravities}. \textit{Class. Quantum Grav.} \textbf{2006}, \emph{23}, 5799.

\bibitem{guha} Guha, P.; Harikumar, E.; Zuhair, N.S. \textit{MICZ-Kepler systems in noncommutative space and duality of force laws}. \textit{Int. J. Mod. Phys.} \textbf{2014}, \emph{29}, 1450187.

\bibitem{hari2} Harikumar, E.; Kapoor, A.K. \textit{Newton's Equation on the $\kappa$-Spacetime and the Kepler Problem}. \textit{Mod. Phys. Lett.} \textbf{2010}, \emph{25}, 2991.

\bibitem{madfr} Harikumar, E.; Panja, S.K.; Rajagopal, V. \textit{Maximal acceleration in a Lorentz invariant non-commutative space-time}. \textit{Eur. Phys. J. Plus} \textbf{2022}, \emph{137}, 966.

\bibitem{kappa1} Lukierski, J.; Nowicki, A.; Ruegg, H.; Tolstoy, V.N.
 \textit{q-deformation of Poincaré algebra}.
\textit{Phys. Lett.} \textbf{1991}, \emph{264}, 331.

\bibitem{dimitrijevic}
Dimitrijevic, M.; Jonke, L.; Moller, L.; Tsouchnika, E.; Wess, J.; Wohlgenannt, M.  \textit{Deformed field theory on $\kappa$
-spacetime}. \textit{Eur. Phys. J.} \textbf{2003}, \emph{31}, 129.

\bibitem{dasz}
Daszkiewicz, M.; Lukierski, J.; Woronowicz, M.  \textit{Towards quantum noncommutative 
$\kappa$
-deformed field theory}.  \textit{Phys. Rev.} \textbf{2008}, \emph{77}, 105007.

\bibitem{mel1}Meljanac, S.; Stojic, M. \textit{New realizations of Lie algebra kappa-deformed Euclidean space}. \textit{Eur. Phys. J.} \textbf{2006}, \emph{47}, 531.

\bibitem{carlson} 
Carlson, C.E.; Carone, C.D.; Zobin, N. \textit{Noncommutative gauge theory without Lorentz violation}. \textit{Phys. Rev.} \textbf{2002}, \emph{66}, 075001.

\bibitem{amorim} 
Amorim, R. \textit{Dynamical symmetries in noncommutative theories}. \textit{Phys. Rev.} \textbf{2008}, \emph{78}, 105003.

\bibitem{gupta1} Gupta, K.S.; Harikumar, E.; Juri\'c, T.; Meljanac, S.; Samsarov, A.  \textit{Effects of Noncommutativity on the Black Hole Entropy}. \textit{Adv. High Energy Phys.}  \textbf{2014}, \emph{2014}, 139172.

\bibitem{gupta2} Gupta, K.S.; Harikumar, E.; Juri\'c, T.; Meljanac, S.; Samsarov, A.  \textit{Noncommutative scalar quasinormal modes and quantization of entropy of a BTZ black hole}. \textit{J. High Energy Phys.} \textbf{2015}, \emph{9}, 025.

\bibitem{gupta3} Gupta, K.S.; Juri\'c, T.; Samsarov, A. \textit{Noncommutative duality and fermionic quasinormal modes of the BTZ black hole}. \textit{J. High Energy Phys.} \textbf{2017}, \emph{6}, 107.

\bibitem{gupta4} Digal, S.; Govindarajan, T.R.; Gupta, K.S.; Martin, X. \textit{Phase structure of fuzzy black holes}. \textit{J. High Energy Phys.} \textbf{2012}, \emph{1}, 027.

\bibitem{gupta5} Gupta, K.S.; Juri\'c, T.; Samsarov, A.; Smoli\'c, I.  \textit{Noncommutativity and logarithmic correction to the black hole entropy}. \textit{J. High Energy Phys.} \textbf{2023}, \emph{2}, 060.

\bibitem{paolo} Aschieri, P.; Borowiec, A.; Pacho, A. \textit{Dispersion relations in $\kappa$-noncommutative cosmology}. \textit{J. Cosmol. Astropart. Phys.} \textbf{2021}, \emph{4}, 025.

\bibitem{jain} Kothari, R.; Rath, P.K.; Jain, P. \textit{Cosmological power spectrum in a noncommutative spacetime}. \textit{Phys. Rev.} \textbf{2016}, \emph{94}, 063531.
\bibitem{bani} Kalita, S.; Govindarajan, T.R.; Mukhopadhyay, B. \textit{Super-Chandrasekhar limiting mass white dwarfs as emergent phenomena of noncommutative squashed fuzzy spheres}. \textit{Int. J. Mod. Phys.} \textbf{2021}, \emph{30}, 2150101.
\bibitem{ohl}  Ohl, T.; Schenkel, A. \textit{Cosmological and black hole spacetimes in twisted noncommutative gravity}. \textit{J. High Energy Phys.} \textbf{2009}, \emph{0910}, 052.
\bibitem{mel2}
Meljanac, S.; Kresic-Juric, S.; Stojic, M. \textit{Covariant realizations of kappa-deformed space}. \textit{Eur. Phys. J.} \textbf{2007}, \emph{51}, 229.
\bibitem{mel3}
Meljanac, S.; Samsarov, A.; Stojic, M.; Gupta, K.S. \textit{$\kappa$-Minkowski spacetime and the star product realizations}. \textit{Eur. Phys. J.} \textbf{2008}, \emph{53}, 295.
\bibitem{hari1a} Harikumar,
 E.; Juric, T.; Meljanac, S. \textit{Electrodynamics on $\kappa$-Minkowski space-time}. \textit{Phys. Rev.} \textbf{2011}, \emph{84}, 085020.
 \bibitem{hari1b}Harikumar, E.; Juric, T.; Meljanac, S. \textit{Geodesic equation in $\kappa$-Minkowski spacetime}.  \textit{ Phys. Rev.} \textbf{2012}, \emph{86}, 045002.
 
\bibitem{vishnu1} Harikumar, E.; Lakkaraju, L.G.C.; Rajagopal, V. \textit{Emergence of maximal acceleration from non-commutativity of spacetime}. \textit{Mod. Phys. Lett.} \textbf{2021}, \emph{36}, 2150069.
\bibitem{Mukanov} Mukhanov, V. \textit{Physical Foundations of Cosmology}; Cambridge University Press: {Cambridge, UK,} 2005.
\bibitem{Steven} Weinberg, S. \textit{Cosmology}; Oxford University Press: {Oxford, UK},  2008.
\bibitem{hu1} {Weinberg,} 
 D.H.; Mortonson, M.J.; Eienstein, D.J.; Hirata, C.; Riess, A.G.; Rozo, E. \textit{Observational probes of cosmic acceleration}. \textit{Phys.
Rep.} \textbf{2013}, \emph{530}, 87.
\bibitem{hu2}{Hu}, J.P.; Wang, F.Y. \textit{Hubble Tension: The Evidence of New Physics}. \textit{Universe} \textbf{2023}, \emph{9}, 94.
\bibitem{planc} Planck Collaboration. Planck 2013 results. XVI. Cosmological parameters. \textit{Astron. Astrophys.} \textbf{2014}, \emph{571},  A16.
\bibitem{ben}  Bennett, C.L.; Larson, D.; Weiland, J.L.; Jarosik, N.; Hinshaw, G.; Odegard, N.; Smith, K.M.; Hill, R.S.; Gold, B.; Halpern, M.;~et~al. \textit{NINE-YEAR WILKINSON MICROWAVE ANISOTROPY PROBE (WMAP) OBSERVATIONS: FINAL MAPS AND RESULTS}. \textit{Astrophys. J. Suppl. Ser.} \textbf{2013}, \emph{208}, 20.
\bibitem{dhawan1}  Dhawan,
 S.; Leibundgut, S.W.J.A.B. \textit{Measuring the Hubble constant with Type Ia supernovae as near-infrared standard candles}. \textit{A\&A} \textbf{2018}, \emph{609}, A72.
 \bibitem{dhawan2}{De} Jaeger, T.; Stahl, B.E.; Zheng, W.; Filippenko, A.V.; Riess, A.G.; Galbany, L. \textit{A measurement of the Hubble constant from Type II supernovae}. \textit{Monthly Not. R. Astron. Soc.} \textbf{2020}, \emph{496}, 3.
\bibitem{hagh} Haghighat, M. \textit{Bounds on the parameter of noncommutativity from supernova SN1987A}. \textit{Phys. Rev.} \textbf{2009}, \emph{79}, 025011. 
\bibitem{Joby}  Joby, P.K.; Chingangbam, P.; Das, S. \textit{Constraint on noncommutative spacetime from PLANCK data}. \textit{Phys. Rev.} \textbf{2015}, \emph{91}, 083503. 
 
\end{thebibliography}
\end{document}